\documentstyle[preprint,aps]{revtex}
\renewcommand{\Re}{\mbox{Re}}
\renewcommand{\Im}{\mbox{Im}}

\newcommand{\eqn}[1]{Eq~(\ref{#1})}

\begin{document}
\preprint{\vbox{\hbox{}}}
\draft
\title{Ruling out the Weinberg Model of Spontaneous CP Violation}
\author{$^{1,4}$Darwin Chang, $^{2,3}$Xiao-Gang He, and $^3$Bruce H.J.
McKellar}
\address{$^1$ Department of Physics, National Tsinhua University,
Hsinchu, 300, Taiwan.\\
 $^2$ Department of Physics, National Taiwan University,
Taipei, 10617,Taiwan.\\
$^3$School of Physics,
University of Melbourne,
Parkville, Vic 3052, Australia.\\ and \\
$^4$SLAC, Stanford, CA 94309, USA.
}
\date{August  1999, revised August 2000}

\maketitle

\begin{abstract}
There have been many declarations of the death of the Weinberg model
of spontaneous CP violation.
Previous studies, before the recent measurements
of $\epsilon'/\epsilon$, indicated that the
model could not accommodate the experimental values on $\epsilon$ in
$K^0-\bar K^0$ mixing, the neutron electric dipole moment (EDM),
the branching ratio of $b\to s \gamma$ and the upper limit on
$\epsilon'/\epsilon$.
We point out that these
studies were based on optimistic estimates of the uncertainties in the
calculations and that when more realistic estimates of these
errors are used
the Weinberg model
cannot be conclusively ruled out from these considerations alone.

Here we use these realistic error estimates  to analyze
the present
situation of the Weinberg model.  The latest results
from Belle and BaBar
on $\sin 2 \beta$ allow the small values of this parameter which
occur naturally
in the Weinberg model.  However, in this model, the recently measured
value of $\Re(\epsilon'/\epsilon) = (1.92\pm 0.25)\times 10^{-3}$ cannot be
made
compatible with the branching ratio $B(b\to s \gamma) =
(3.15\pm0.54)\times 10^{-4}$.
As a result we conclude that the Weinberg model is
now confidently and conservatively ruled out.
\end{abstract}

\section{Introduction}
The origin of CP violation remains one of the outstanding problems of
modern
particle physics. Although the Standard Model (SM) of CP violation
based on the
Kobayashi-Maskawa (KM) mechanism
is consistent\cite{1} with observations of CP violation
in $K_S$ and $K_L$ mixing\cite{1a} and
in $K_{S,L} \to \pi\pi$ decay amplitudes\cite{1b},
there are intriguing hints,  from  consideration of baryon
asymmetry of the universe\cite{2}, that other sources of CP violation
may exist.  These non-standard sources of CP violation could occur as
well as, or instead of, the SM source.

Models based on additional Higgs bosons\cite{3,4}
provide alternatives which explain the existing laboratory
data\cite{5} and produce the large CP violation
required for baryon asymmetry\cite{2}.
Such models also allow CP symmetry to be broken
spontaneously\cite{4,branco} and
therefore give an interesting explanation of the origin for CP
violation.  The minimal model of this type satisfying the
requirement of vanishing tree level flavour changing neutral current
(``natural flavor conservation'') is a model of spontaneous CP
violation with three Higgs doublets proposed by Weinberg\cite{4} and
refined by Branco\cite{branco}.  We shall refer to it as Weinberg
model from here on.

It has frequently been claimed that the Weinberg model is in conflict
with the data on %two or more of
the following:
\begin{itemize}
	\item the value of $\sin 2 \beta$.
    	\item the ratio $\epsilon'/\epsilon$,
	\item the  $\epsilon$ parameter in $K^0-\bar K^0$ mixing,
	\item the neutron electric dipole moment (EDM),
	\item the branching ratio of $b\to s \gamma$,
\end{itemize}

First of all, the Weinberg model predicts small values (less than
$0.05$) for the
parameter $\sin 2 {\beta}$ which here we define as the parameter that
characterizes the CP violation
in $B \to J/\psi K_S$ decays\cite{13,14}.
The present results\cite{belle} from Belle
($0.45^{+0.43}_{-0.44}(\mbox{stat}) ^{+0.07}_{-0.09}(\mbox{sys})$) and
BaBar ($0.12 \pm 0.37 \pm 0.09$) on
$\sin 2 \beta$ allow such small values of this parameter, although
the
earlier combined
result from ALEPH, OPAL and CDF($0.91\pm 0.35$)\cite{forty}
favoured larger values.
Therefore these considerations do not rule out the Weinberg model,
and to determine whether or not the model is consistent with present
day data we have to turn to the other observables.

In early discussions of CP violation in the neutral kaon system, it
was assumed that only the short-distance contributions from the
CP violating $\Delta S = 2$ box diagrams, due to either two charged
Higgs particles, or one charged Higgs plus one $W$ exchange,
are responsible for the measured value
of $\epsilon$, then, since the charged Higgs couplings to light
fermions are proportional to the fermion masses a very large CP
violating coupling is required to fit the data.  If this same CP
violating parameter is then used for the calculation of
$\epsilon'/\epsilon$, the contribution is much larger than the
experimental value\cite{8,9}.  It was later shown that
there are important long distance contributions to $\epsilon$ due to the Higgs
induced $\Delta s = 1$ CP violating operators, and that if these
are taken into account, the model can be made consistent with the
observed $\epsilon$ and $\epsilon'/\epsilon$~\cite{10,11}.

Previous studies have claimed that although the Weinberg model is
consistent with CP violation in kaon system, it has problems with the
neutron electric dipole moment (EDM)\cite{12} and the branching ratio
for $b\to s\gamma$\cite{13,14}.  These analyses have relied on
optimistic estimates of the accuracy of the calculations of relevant
hadronic matrix elements\cite{16a}.

After briefly describing the Weinberg model of CP
violation, we review the current experimental and theoretical status
of each of these observables individually, and then discuss the
constraints placed on the Weinberg model by the total ensemble of
data.  We find that it is possible to rule out this model
definitively only if one imposes simultaneously the new constraints
from the recently
measured value of $\Re(\epsilon'/\epsilon) = (1.92\pm 0.25)\times
10^{-3}$\cite{1b} and the branching ratio $B(b\to s \gamma) =
(3.15\pm0.54)\times 10^{-4}$\cite{cleogamma}.

Therefore, because of the stronger constraints imposed by recent data,
one can now declare that the Weinberg model of spontaneous CP
violation is ruled out in spite of the relatively
large hadronic uncertainty.

\section{The Weinberg model of CP violation}

In the Weinberg model, three Higgs doublets are introduced.  The
spontaneous breakdown of gauge symmetry then induces massive Higgs
eigenstates for the charged Higgs particles, and introduces a mixing
matrix specifying the interaction eigenstates of the Higgs doublets in
term of the mass eigenstates.  Being a $3 \times 3$ mixing matrix
between charged particles, the matrix contains exactly one irreducible
complex phase, thus inducing CP violation.  After spontaneous symmetry
breaking, there are two physical charged and five neutral Higgs
particles.  As CP violation in flavour changing processes in this
model is dominated by exchange of charged Higgs particles, we
concentrate our attentioin on this contribution.  The interaction
Lagrangian for the coupling of the two charged Higgs ($H_1^+$ and
$H_2^+$) to fermions\cite{6} can be written as
\begin{eqnarray}
L = 2^{3/4}G_F^{1/2}\bar U [V_{KM}M_D(\alpha_1H_1^+ + \alpha_2 H^+)R
+M_UV_{KM}(\beta_1 H^+_1 +\beta_2 H_2^+)L]D + H.C.\;, \label{eq:Higgs}
\end{eqnarray}
where $R(L) = (1\pm\gamma_5)/2$, and $M_{U,D}$ are the diagonal up and
down quark mass matrices.  The parameters $\alpha_i$ and $\beta_i$,
which satisfy ${\Im}(\alpha_1\beta^*_1) = -
{\Im}(\alpha_2\beta_2^*)$, are obtained from the diagonalisation of
the charged Higgs mass matrix .
The KM matrix elements $V_{ij}$ can be made
all real at tree level as a consequence of spontaneous CP violation.

\section{$\sin 2 \beta$ in the Weinberg model}
It is well known that CP violation in B decays will eventually
provide crucial constraints on models of CP violation when  sufficient
data is available.  This is especially true
of the gold-plated mode $B\to J/\psi
K_S$.

In the Weinberg model, CP violating contributions to the decay
amplitudes and to $B^0-\bar B^0$ are both proportional to
$\Im(\alpha_1 \beta^*_1)$.
The Higgs contributions to the decay amplitudes
are suppressed, relative to the SM
contributions, by additional factors of $m_cm_b/m_H^2$, while the
mixing is supressed by a factor of $m_b^2/m_H^2$.
These suppression factors lead to small CP violating phases and result
in a very small value for $\sin 2{\beta}$\cite{13,14}, $|\sin
2{\beta}| <0.05$.

The ALEPH, OPAL and CDF data reported in 1999 gave $\sin 2
{\beta} =
0.91\pm 0.35$\cite{forty} which is in conflict with the above limit
at the $2\sigma$ level.  However at ICHEP2000, Belle and BaBar
reported
preliminary results\cite{belle}, which when averaged with the above,
give $\sin 2 {\beta} = 0.49 \pm 0.45$, consistent with the
above limit
at the 75\% level.  Thus the present $\sin2{\beta}$
measurements do not rule out the Weinberg model.

\section{ $\epsilon'/\epsilon$ in the Weinberg model}

The dominant contribution to $\epsilon'/\epsilon$ in the Weinberg
model is from the flavor changing gluonic dipole interaction given
by\cite{13}
\begin{eqnarray}
&&H(sdg) = ig_s \tilde f m_s \bar s \sigma_{\mu\nu} \lambda^a
G_a^{\mu\nu} (1-\gamma_5)
d,\nonumber\\
&&\tilde f = {G_F\over \sqrt{2}} {1\over 16\pi^2}
V_{is}V_{id}\Im(\alpha_1^*\beta_1)
(F_3(m^2_i/m^2_{H_1})-F_3(m^2_i/m^2_{H_2}))\eta_g,\nonumber\\
&&F_3(x)= {1\over 2} {x\over (1-x)^3}[ - {3\over 2} + 2 x -{1\over 2}
x^2
- \ln x], \label{eq:sgd}
\end{eqnarray}
where $i$ is summed over $u,\;c,\;t$ and $\eta_g = (\alpha_s(m_H)/
\alpha_s(\mu))^{14/(33-2n_f)}$ is the one loop QCD correction
factor\cite{qcd} in which $n_f$ being the number of quark with mass
less than $\mu$.  To obtain this correction factor we will use one
loop running for $\alpha_s$  with the starting value $\alpha_s(m_Z) =
0.119$.
The contribution to $\epsilon'/\epsilon$ is dominated by the lightest
charged Higgs exchange.  In our later discussions, we will assume
$H_1^+$ is the lighter one and the other is very heavy and its effects
can be neglected.

Theoretical analyses for $\epsilon'/\epsilon$ are conventionally
carried out in terms of the isospin amplitude $A_I$ for $K\to
\pi\pi$.
Expressing $\epsilon'/\epsilon$ in terms of $A_I$, one obtains
\begin{eqnarray}
\Re\left ( {\epsilon'\over \epsilon}\right ) \approx
{\omega\over \sqrt{2} |\epsilon|} \left ({\Im A_2\over \Re A_2} -
{\Im
A_0\over \Re A_0}\right ),
\end{eqnarray}
where $\omega = \Re A_2/\Re A_0 \approx 1/22.2$.

The dominant gluon dipole operator \eqn{eq:sgd} generates a non-zero
value only for $A_0$.  Calculating the decay amplitudes is our most
difficult task, because of our poor understanding of the
strong interaction at low energies.  Theoretical calculations for the
real part of the amplitudes can be easily off by a factor of two to
three.  For this reason we use the experimental value for $\Re A_0=
33.3\times 10^{-8}$ GeV$^{-2}$ to minimize the error in the
calculation of $\epsilon'/\epsilon$.  But we still have to calculate
$\Im A_0$.  This requires the evaluation of the matrix element
$<(\pi\pi)_0|{\cal O}|K>$.  Here $(\pi\pi)_0$ indicates the isospin
$I = 0$
component and ${\cal O} = g_s m_s \bar s \lambda^a
\sigma_{\mu\nu}G_a^{\mu\nu} (1-
\gamma_5)
d$.

A naive PCAC calculation\cite{15,16} gives
\begin{equation}
\langle (\pi\pi)_0|{\cal O}|K \rangle = -2\sqrt{3/2}(m_0^2
m_s/(m_u+m_s))(m_K^2 f_K/f_\pi)  \label{kpipi}
\end{equation}
with $m_0^2 \approx 1 \mbox{GeV}^2$.
 A bag model calculation\cite{9} of
\begin{equation}
A_{K\pi} = <\pi^0|\bar s \lambda^a \sigma_{\mu\nu}G^{\mu\nu}_a(1-\gamma_5) d|K>
\label{akpi}
\end{equation}
gives
\[
A_{K\pi}\approx 0.4 \mbox{GeV}^3,
\]
and with the use of current algebra this\cite{11} gives a value for
$\langle (\pi\pi)_0|{\cal O}|K \rangle$ similar to that of
\eqn{kpipi}.

It was later realized that the above result is incorrect because an
important ``tadpole'' contribution due to the K-vacuum transition
caused by the same operator had been neglected.  This contribution
cancels the above PCAC result exactly\cite{11,15}.
In a chiral
perturbation theory approach, this means that the leading order
contribution vanishes as expected from the Feinberg-Kabir-Weinberg
Theorem\cite{11}.  A non-zero value for $<(\pi\pi)_0|{\cal O}|K>$ can only be
generated at $p^4$ order in chiral perturbation theory, and can be
estimated to be\cite{16}
\begin{eqnarray}
<(\pi\pi)_0|{\cal O}|K> = -11\sqrt{{3\over 2}}{m_s\over m_s +m_d}
{f^2_K\over
f^3_\pi} m_K^2 m_\pi^2 B_0,
\end{eqnarray}
where $B_0$ is a fudge factor representing the potential
uncertainty in the above estimate.
We assume that $B_0$ is of order $1$.

Using this matrix element,
we obtain
\begin{eqnarray}
\Re\left ( {\epsilon'\over \epsilon}\right ) &=& {\omega\over
\sqrt{2}|\epsilon| Re A_0} 11\sqrt{{3\over 2}} {m_s\over m_s +m_d}
{f_K^2\over f_\pi^3} m_K^2 m_\pi^2\tilde f B_0 (1-\Omega_{\eta+\eta'})
\nonumber\\
&=&  1.7 \times 10^{7} (\mbox{GeV}^2) \tilde f B_0, \label{epe}
\end{eqnarray}
where the numerical value follows from  the experimental values for
$\epsilon$ and $Re A_0$, and the isospin breaking correction
factor $\Omega_{\eta+\eta'} = 0.25$ given in ref~\cite{iso}

To produce the recently observed value for $\epsilon'/\epsilon$ within
$3\sigma$, $\tilde f B_0$ has to be in the range $+(0.69 \sim
1.57)\times 10^{-10}$ GeV$^{-2}$.  For a given Higgs mass, the CP
violating parameter ${\Im}(\alpha_1\beta_1^*)$ is determined by the
value of $\tilde{f}$.  Only $\tilde f B_0$ is determined by the data.
Remember that the different leading order contributions cancel each
other.  However, numerically the value obtained from \eqn{epe} with
$B_0 =1$ is not much smaller than the individual leading terms before
cancellations, suggesting that $B_{0} = 1$ is the maximum value of
$B_{0}$ and hence that the corresponding low value for $\tilde{f}$,
$\tilde{f} = 0.69\times 10^{-10}$ GeV$^{-2}$, represents its probable
lower bound.  Nevertheless, we conservatively allow $B_0$ to vary from
0.5 to 2 in our estimate to account for possible
uncertainties\cite{hg}.  The most conservative range for $\tilde f$ is
then $(0.35 \sim 3.1) \times 10^{-10}$ GeV$^{-2}$.  Thus $\tilde f$
smaller than $0.35\times 10^{-10}$ GeV$^{-2}$ in magnitude is unlikely
to generate $\epsilon'/\epsilon$ as large as observed.  Note that
$\tilde{f}$ is positive because $B_0$, calculated in Ref\cite{16}, is
positive.

\section{$\epsilon$ in the Weinberg model}
A successful model for CP violation must to be able to produce the
experimental value for $\epsilon$.  In this model the short distance
$\Delta S =2$ interaction gives too small a value for $\epsilon$, and
the dominant contribution actually comes from long distance effects,
which in turn are generated by CP violation due to the gluonic dipole
interaction.  Following Ref.\cite{11} we assume the contribution to
$\epsilon$ is from  $\pi,\;\eta,\;\eta'$ poles with one CP
conserving and one CP violating K to $\pi,\;\eta,\;\eta'$ transition.
One has\cite{11}
\begin{eqnarray}
&&|\epsilon| = {\tilde f \kappa g_s m_s A_{K \pi } H_{\pi K} \over
\sqrt{2} m_K \Delta m_{L-S} (m_K^2 - m_\pi^2)},\nonumber\\
&&\kappa= 1+ {m_K^2-m_\pi^2\over m_K^2-m_\eta^2}
\left[\sqrt{{1\over 3}} (1+\delta) \cos\theta + 2\sqrt{{2\over 3}}
\rho \sin\theta\right]^2\nonumber\\
&&\;\;\;\; + {m_K^2-m_\pi^2\over m_K^2-m_{\eta'}^2}
\left[ \sqrt{{1\over 3}}(1+\delta) \sin\theta
 - 2\sqrt{{2\over 3}}\rho \cos\theta\right]^2,
\end{eqnarray}
where $\Delta m_{L-S}$ is the mass difference of the long and short
lived neutral kaons, $\theta$ is the $\eta-\eta'$ mixing angle, and
$\delta$ and $\rho$ parameterize SU(3) and U(3) breaking effects,
respectively.  In the SU(3) limit, $\delta = 0$; in the U(3) limit,
$\rho = 1$.  $H_{\pi K}$ is the CP conserving $\Delta S = 1$, $K-\pi$
transition amplitude which is determined from current algebra to
be\cite{11} $H_{\pi K} = 2.578\times 10^{-8}$ GeV$^2$.  $\tilde H_{\pi
K} = \tilde f g_s m_s A_{K \pi}$ is the CP violating $K-\pi$
transition amplitude due to the gluonic dipole interaction, expressed
in terms
of the matrix element $A_{K \pi}$ of \eqn{akpi}.
Here the QCD coupling constant $g_s$ is evaluated at the Kaon scale
and is not well determined.
Following Ref. \cite{weinberg}, we use $g_s=4\pi/\sqrt{6}$ in the
matrix element calculation.  For $m_s$, one should take the values
used in
conjunction with the models used to calculate relevant matrix
elements, for
example in the bag model calculation it is in the range of 0.3 to 0.5
GeV\cite{bag}.

The parameters, $\theta$, $\delta$, $\rho$, and the theoretical
calculation of $A_{K \pi }$, introduce uncertainties.  At present,
there are two possible values $-11^o$ and $-22^o$\cite{17} for the
mixing angle $\theta$.  The SU(3) breaking parameter $\delta$ is
theoretically estimated\cite{18}.  The best fit\cite{19} $K_L\to
\gamma\gamma$ and $K\to \pi\pi \gamma$ gives a U(3) breaking parameter
$\rho = 0.78$.  However, $\rho$ in the range of $0.7 \sim 1.3$ is not
ruled out.  Using $\theta = -22^o$, $\delta = 0.17$ and $\rho =0.78$,
one obtains $\kappa = 0.2$.  Most of the previous calculations used
this value for $\kappa$.
However, the value for $\kappa$ is very sensitive to the
specific values of the parameters involved, for example, with $\theta
= -11^o (-22^o)$, $\rho = 1.3$, and $\delta = 0.17(0.0)$, $\kappa$ is
approximately -0.9 (-0.95).  The magnitude of $\kappa$ can change by a
factor of four or five.  Within the allowed parameter space, $\kappa$
can vary between 0.2 to -1.0.  We note that the sign of $\kappa$
changes in the allowed range of parameters which implies that the
relative sign of $\epsilon$ and $\epsilon'$ can change.

The uncertainty in the value $A_{\pi K}$ is also quite large.  A bag
model calculation gives $A_{\pi K} = 0.4 GeV^3$\cite{9}.
Major sources of uncertainty include the
determination of the numerical values of $\alpha_s$ at the kaon decay
scale, of the bag radius $R$ and of the strange quark mass $m_s$ in
bag model \cite{bag}.  A factor of two to three times increase in
$A_{\pi K}$ is not ruled out.
In view of
these uncertainties, we consider $(\kappa, A_{K\pi})$
in the rectangle with corners $(0.2,\;0.4 \mbox{GeV}^3)$ and
$(-1.0,\;1.2\mbox{GeV}^3)$
to be allowed by present
experimental and theoretical estimates.  Of the two extreme
values are, set a) $(0.2, 0.4 \mbox{GeV}^3)$ and set b) $(-1.0, 1.2
\mbox{GeV}^3)$,  set a) is the mostly used one in the literature,
while set b) represents the most conservative values for $\kappa$ and
$A_{K\pi}$.

We find that if the parameters of set a) are used to fit $\epsilon$,
the parameter $\tilde f$ is determined to be $2.56\times 10^{-10}$
GeV$^{-2}$.  However, if set b) is used, $\tilde f$ can be negative
and as small as $0.17\times 10^{-10}$ GeV$^{-2}$ in magnitude.  There
are solutions for $\epsilon$ with $\tilde f$ in the ranges $(0.85 \sim
2.56)\times 10^{-10}$ GeV$^{-2}$ and also with $\tilde f$ near
$-0.17\times 10^{-10}$ GeV$^{-2}$.  The allowed range of $\tilde f$
associated with $\epsilon$ is thus quite large, and it has a large
overlap with that determined from $\epsilon'/\epsilon$ for positive
$\tilde f$.

\section{The neutron electric dipole moment}
The experimental bound on the neutron EDM, $d_{n}$, has been used to
provide restrictions on the model, and has been claimed to rule it
out\cite{12}.  The neutron EDM can be generated by the exchange of
neutral and charged Higgs particles\cite{21,22,23,24}.  It is not
impossible
that these contributions may cancel each other and result in a very
small neutron EDM. Here we will not entertain this possibility.  We
will instead single out the variously potentially large valence quark
contributions and require that each of them satisfies the experimental
constraints.

The contribution of charged Higgs exchange  to the neutron EDM
is well constrained by fixing the CP violating parameter
${\Im}(\alpha_1\beta_1^*)$ to fit $\epsilon'/\epsilon$ and
$\epsilon$.
The contributions from neutral Higgs exchange are much less
constrained.  Even in the charged Higgs case we need to use the
theoretical expression for $\tilde f$ to extract
${\Im}(\alpha_1\beta_1^*)$, and this introduces a sensitivity to the
values of the KM elements because the internal charm and top
contributions are comparable and can add constructively or
destructively depending on the relative sign of combinations of the KM
matrix elements.  This also introduces uncertainties in the
calculations.  The case where the contributions tend to cancel will
result in a large $\Im(\alpha_1 \beta^*_1)$ and lead to difficulties
with other data as discussed below.  We will use values of the KM
matrix elements within the errors given in Ref.\cite{1a} such that
terms contribute constructively.
Specifically, for later discussions, we use $V_{ud} =
0.9741$, $V_{us}=0.221$, $V_{cd}=-0.220$, $V_{cs} = 0.9740$ and
$V_{ts}=-0.040$.

The charged Higgs boson contribution to the
neutron EDM is strongly restricted. The dominant term comes
from the down quark EDM. Using the valence quark model, we have
\begin{eqnarray}
&&d_n \approx {4\over 3} d_d =  {8\over 3}e m_d
\tilde f {V_{cd}^2F_2(m^2_c/m_{H_1}^2) +\eta_\gamma V_{td}^2
F_2(m_t^2/m^2_{H_1})
\over V_{cd}V_{cs}F_3(m_c^2/m_{H_1}^2)+\eta_g
V_{td}V_{ts}F_3(m_t^2/m_{H_1}^2)},
\nonumber\\
&&F_2(x)= - {x\over 6(1-x)^3} [ (3-5x)(1-x) + (4-6x)\ln x].
\end{eqnarray}
Here we have neglected the small QCD correction to the electric dipole
operator from the gluonic dipole operator induced by operator
mixing.  The
leading QCD correction factor for the  electric dipole operator is
given by\cite{qcd} $\eta_\gamma =
[\alpha_s(m_W)/\alpha_s(\mu)]^{16/(33 - 2 n_{f})}$.

Using $\tilde f =(0.35\sim 3.1) \times 10^{-10}$ GeV$^{-2}$ determined
from $\epsilon'/\epsilon$, and alowing the lightest charged Higgs mass
to range from its lower bound around 70 GeV to several hundred GeV, we
estimate the charged Higgs contribution to the neutron EDM as $ (0.25
\sim 3.5)\times 10^{-24} (m_d/300{\rm MeV}) {\rm e cm}$.
Note
that $d_{n}$ is proportional to the light quark mass.  This introduces
a further uncertainty because it is not clear whether the current or
the constituent mass should be used.  There are also other
uncertainties due to the off shell nature of the quarks\cite{bruce}.

The neutral Higgs boson exchange gives a contribution
which is not well determined.
Even the sign of the contribution is unknown.
Since it is not related to the other parameters we have
introduced, and since a wide range of the parameter $\tilde f$ is
still
allowed, we will not consider the possible contributions from
exchange
of neutral Higgs bosons in our estimates.

\section{$b \to s \gamma$ in the Weinberg model}

The CP conserving process $b \to s \gamma$ can place constraints on
the CP violating parameters of the model\cite{13,14}, because the CP
violating amplitudes contribute to the total rate. In the Weinberg
model, although $ds\gamma$ interaction is constrained to be small, the
corresponding $bs\gamma$ interaction is enhanced by a factor of $
(\sim {m^2_t m_b/ m_c^2 m_s}) ({V_{tb} V_{ts}/ V_{cs}V_{cd}}) \sim
10^{5}$.  Due to this enhancement factor, the predicted branching
ratio of $b\to s \gamma$ may be in conflict with experimental data.
Using the leading log result and normalizing the branching ratio due
to charged Higgs contribution to the SM one, we have\cite{25}
\begin{eqnarray}
&&Br(b\to s\gamma) =
7.1\times 10^{-4}[(0.313 + 0.273 r_1)^2 + (0.273r_2)^2],\nonumber\\
&&r_1 = 1 + {|\beta_1|^2F_1(m_t^2/m_{H_1}^2)/3 - {\rm Re}(\alpha_1
\beta_1^*)F_2(m_t^2/m_{H_1}^2)\over F_1(m_t^2/m_W^2)},
\nonumber\\
&&r_2 = - {{\rm Im}(\alpha_1 \beta_1^*)
F_2(m_t^2/m_{H_1}^2)\over F_1(m_t^2/m_W^2)},\nonumber\\
&&F_1(x)= {x\over 12(1-x)^4} [(7-5x-8x^2)(1-x) + x(12-18x)\ln x].
\end{eqnarray}
In the above we have neglected the small contribution from the gluonic
$bsg$ interaction.  CP conserving amplitudes generate the first term
in the brackets, and $r_{1}$ contains all of the contributions
dependent on $m_{t}$.  There are both SM and Weinberg model
contributions, and there is a region in parameter space where that the
CP conserving contributions of the SM and of charged Higgs exchange
mutually cancel.  The CP violating amplitudes generate $r_{2}$, which
contributes significantly.  The branching ratio increases with Higgs
mass for fixed $\tilde f$.

The experimental branching ratio\cite{cleogamma}, $B(b\to s\gamma) =
(3.15\pm0.54)\times 10^{-4}$, has recently been confirmed by
Belle\cite{belle}.  For the 95\% c.l. upper bound $4.5\times 10^{-4}$
for $b\to s\gamma$\cite{cleogamma}, we find that there are solutions
with $|\tilde f| \le  0.17\times 10^{-10}$ GeV$^{-2}$ for the charged
Higgs mass greater than 70 GeV. For larger values of $m_{H}$, tighter
constraints are placed on $|\tilde f|$.  Cutting the photon energy
$E_\gamma$ to be larger than $2.1$ GeV to ensure that the contribution
is indeed due to the penguin diagram contribution considered here, the
central value of the branching ratio is reduced slightly to
$2.97\times 10^{-4}$.  The allowed range, at fixed $\tilde f$, for the
Higgs mass is restricted at the upper end by this reduction.  For
example, $|\tilde f| = 0.17\times 10^{-10}$ GeV$^{-2}$ is consistent
with the reduced branching ratio for $70\mbox{GeV} \le m_{H} \le
110\mbox{GeV}$.  There is a region in which the constraints on $\tilde
f$ from $b\to s\gamma$,
$\epsilon$ and the neutron EDM are consistent.  But it is not
possible
to simultaneously satisfy the constraints from $b \to s \gamma$ and
from $\epsilon'/\epsilon$.

\section{Discussion}

No constraint is yet placed on the model by the results
for $\sin 2 {\beta}$ from ALEPH, OPAL, CDF, BaBar or Belle.

If the current quark mass $m_d \sim 10$ MeV is used, the resulting
value of $d_{n}$ satisfies the experimental limit as long as
$\tilde f < 2.56\times 10^{-10}$ GeV$^{-2}$.  However, if
constituent mass $m_d \sim 300$ MeV is used, the model may be in
trouble.
We know of no convincing argument for preferring one mass
over the other, and therefore conservatively use the current quark
mass to estimate limits.

The values of $\tilde f$ from $\epsilon'/\epsilon$, $\epsilon$ and
$d_{n}$ then have a region of consistency, as do the values
constrained by
$B(b \to s \gamma)$, $\epsilon$ and $d_{n}$.

However there is a definite conflict between the limits on $\tilde f$
from $\epsilon'/\epsilon$ and $B(b \to s \gamma)$.   The latter
requires $|\tilde f| \le 0.17 \times 10^{-10}$ GeV$^{-2}$, and the
former requires
$  0.35 \times 10^{-10}$ GeV$^{-2} \le \tilde f \le 3.1 \times
10^{-10}$ GeV$^{-2}$.
As we
have been careful to make very conservative estimates of the allowed
range of $\tilde f$ (in the hope of finding that there was still a
small region of parameter space in which the model is consistent with
the data), the gap between these allowed regions for $\tilde f$ is
unbridgeable.  Thus we conclude that the Weinberg model is ruled out
by the recent data for $\epsilon'/\epsilon$ and for $B(b \to s
\gamma)$.

One of the attractive features of the version of Weinberg model we
discuss here is that CP
violation is generated spontaneously, rather than being put in by
hand.  If we abandon this attractive feature, and explicit CP
violation is introduced into the Higgs interaction, as in
\eqn{eq:Higgs}, as well as into the W interaction by a phase in the KM
matrix, this new model (which we will call the modified Weinberg
model) is not ruled out.  The constraint from $b\to s\gamma$ requires
that the contribution to $\epsilon'/\epsilon$ from the Higgs
interaction is small, and the main contribution to this CP violating
parameter is just the same as that of the SM. As has been pointed out
in Ref.  \cite{1} there are large uncertainties in the SM calculations
due to our poor understanding of the hadronic matrix elements.  One
can find allowed regions in parameter space in which the experimental
value for $\epsilon'/\epsilon$ is produced in the SM, and thus in the
modified Weinberg model as well.  On the other hand, since the charged
Higgs exchange can also contribute appreciably to $\epsilon$ in this
model and can partially cancel the KM contribution, as a result the KM
phase can have a larger allowed range than that in SM and can easily
accomodate the larger value of $\epsilon'/\epsilon$.  Also, in this model,
the value of $d_{n}$ can still be as large as $3\times
10^{-25}(m_d/300\mbox{MeV}) $e\ cm, very different from the tiny value
of the SM\cite{22}.  And $\sin 2 \beta$ can take the large values
characteristic of the SM.

\acknowledgments This work was supported in part by grants NSC
88-2112-M-002-041, and NSC 89-2112-M-007-010 of the National Science
Council of R.O.C., and in part by the Australian Reserach Council.  DC
wishes to thank the US Department of Energy for partial support while
at SLAC. BMcK thanks the Department of Physics of National Taiwan
University for their hospitality.\\

\end{document}